\documentclass[12pt]{iopart}
\usepackage{txfonts}
\usepackage{ulem}
\usepackage{graphicx}
\usepackage{dcolumn}
\usepackage{bm}
\usepackage{xcolor}

\begin{document}

\title{Pressure-induced enhancement of the superconducting transition temperature in La$_2$O$_2$Bi$_3$AgS$_6$}

\author{Esteban I. Paredes Aulestia$^1$, Xinyou Liu$^1$, Yiu Yung Pang$^1$, Chun Wa So$^2$, Wing Chi Yu$^2$, Swee K. Goh$^1$\thanks{skgoh@cuhk.edu.hk}, Kwing To Lai$^1$\thanks{ktlai@phy.cuhk.edu.hk}}
\address{$^1$Department of Physics, The Chinese University of Hong Kong, Shatin, New Territories, Hong Kong \\ $^2$Department of Physics, City University of Hong Kong, Kowloon, Hong Kong \\}
\ead{skgoh@cuhk.edu.hk, ktlai@phy.cuhk.edu.hk}
\vspace{10pt}
\begin{indented}
\item[]August 2021
\end{indented}

\begin{abstract}
    Charge density wave (CDW) instability is often found in  phase diagrams of superconductors such as cuprates and certain transition-metal dichalcogenides. This proximity to superconductivity triggers the question on whether CDW instability is responsible for the pairing of electrons in these superconductors. However, this issue remains unclear and new systems are desired to provide a better picture. Here, we report the temperature-pressure phase diagram of a recently discovered BiS$_2$ superconductor La$_2$O$_2$Bi$_3$AgS$_6$, which shows a possible CDW transition at $T^*\sim$155 K and a superconducting transition at $T_c\sim$1.0 K at ambient pressure, via electrical resistivity measurements. Upon applying pressure, $T^*$ decreases linearly and extrapolates to 0~K at 3.9~GPa. Meanwhile, $T_c$ is enhanced and reaches maximum value of 4.1~K at 3.1~GPa, forming a superconducting dome in the temperature-pressure phase diagram.
\end{abstract}{

\section{Introduction}

The interplay among different electronic states has in many instances promoted the discovery of exotic states of matter. For example, the emergence of high-$T_c$ superconductivity in the proximity of density wave instabilities in cuprates \cite{Tranquada1995,Fradkin2015,Sachdev2010,DoironLeyraud2012,Sebastian2015,Caprara2017} and iron-based superconductors \cite{Chen2009,Pratt2009,Parker2010,Pratt2011,Luo2012} has been widely studied. Through tuning non-thermal parameters such as doping level and physical pressure, a variety of phase diagrams can be investigated, which help to reveal the key mechanisms for the formation of many exotic states including unconventional superconductivity. \par

An interesting issue in phase diagram studies concerns the role of the charge density wave (CDW) on the appearance of superconductivity. Intricate interaction between CDW and superconductivity can be found in the phase diagrams of several systems, such as cuprate superconductors \cite{Caprara2017},  transition-metal dichalcogenides \cite{Neto2001,Morosan2006,Feng2012,Niu2020}, Kagome metals \cite{Ortiz2020,Du2021,Chen2021}, (Sr,Ca)$_3$(Ir,Rh)$_4$Sn$_{13}$~\cite{Klintberg2012,Goh2015,Yu2015}, Lu(Pt$_{1-x}$Pd$_x$)$_2$In~\cite{Gruner2017}, as well as certain BiS$_2$-based superconductors \cite{Wan2013,Zhai2014,Zhai2014b,Chan2018,Iwasaki2019,Mizuguchi2015a}. However, the precise relationship between CDW and superconductivity is controversial. For instance, in the family of BiS$_2$-based superconductors,  which are structurally analogous to iron-based superconductors, EuBiS$_2$F and Eu$_3$Bi$_2$S$_4$F$_4$ \cite{Zhai2014,Zhai2014b} exhibit the interplay of possible CDW and superconductivity. They are regarded as self-doping systems and superconduct below 0.3 K and 1.5 K for EuBiS$_2$F and Eu$_3$Bi$_2$S$_4$F$_4$, respectively. Both systems show a CDW-like transition at 280 K. Through pressure tuning, the CDW-like transition in EuBiS$_2$F does not change significantly but $T_c$ increases \cite{Guo2015}. In Se-doped Eu$_3$Bi$_2$S$_{4-x}$Se$_x$F$_4$, $T_c$ is enhanced with the suppression of the CDW-like transition \cite{Zhang2015}. This contrasting behaviour makes it difficult to conclude whether the possible CDW phase helps to stabilize the superconducting phase, or compete with superconductivity in BiS$_2$-based systems. \par

The discovery of La$_2$O$_2$Bi$_3$AgS$_6$ \cite{Hijikata2017} in 2017 opens up a new opportunity to study the interplay between CDW and superconductivity in BiS$_2$-based superconductors. Measurements of resistivity on La$_2$O$_2$Bi$_3$AgS$_6$ revealed a superconducting transition at $T_c\approx$ 1.0 K and a possible CDW transition at $T^*\approx$ 155~K \cite{Jha2018}. Through Sn-substitution, the corresponding phase diagram of La$_2$O$_2$Bi$_3$Ag$_{1-x}$Sn$_{x}$S$_{6}$ reported by Jha \textit{et al.} \cite{Jha2019} showed that $T_c$ is enhanced up to $\sim$2.3 K at $x\sim$ 0.4, while $T^*$ is slightly suppressed at $x<$ 0.2 and quickly becomes untraceable at $x>$ 0.2 due to the severe broadening of the anomaly in electrical resistivity associated with $T^*$. In this regard, the relationship between the CDW-like transition and superconductivity in La$_2$O$_2$Bi$_3$AgS$_6$ has not been settled by substitution. \par

In this study, hydrostatic pressure is used as a tuning parameter to study the evolution of $T^*$ and $T_c$ of La$_2$O$_2$Bi$_3$AgS$_6$. Compared with chemical doping, pressure has a lesser impact on the disorder levels of the system, enabling us to track $T^*$ across a larger energy scale. By measuring the temperature dependence of the resistivity of polycrystalline La$_2$O$_2$Bi$_3$AgS$_6$ under pressure up to 7.43~GPa, we present the temperature-pressure phase diagram showing the existence of a superconducting dome centered around the pressure where $T^*$ extrapolates to 0 K. \par
 
\section{Samples and experimental setup} 

Polycrystalline samples of La$_2$O$_2$Bi$_3$AgS$_6$ were synthesized by a solid-state reaction method \cite{Hijikata2017,Jha2018}. The starting materials, La$_2$S$_3$ (99\%), AgO (99.9\%), S (99.98\%), Bi$_2$O$_3$ (99.999\%) and Bi (99.999\%) powder, were well mixed and then pressed into a pellet. The pellet was then inserted into a quartz tube. These procedures were performed in a pure Ar-filled glovebox. The quartz tube was sealed after being evacuated to $<10^{-3}$ mbar using a turbopump. Finally, the reactants inside the sealed quartz tube were annealed at 720 $^{\circ}$C for 15 hr. The annealing process was repeated once after regrinding the pellet. The crystal structure of the samples was examined by x-ray diffraction via Rigaku Smartlab x-ray diffractometer. High-pressure electrical resistivity 
was measured by using a four-point technique in a diamond anvil pressure cell with glycerol as the pressure transmitting medium. The pressure achieved was determined using ruby fluorescence. Low temperature and high magnetic field environments were provided by a Bluefors dilution fridge and a Quantum Design Physical Property Measurement System (PPMS). \par

\begin{figure}[!tb]
    \centering
	\includegraphics[width=8cm]
	{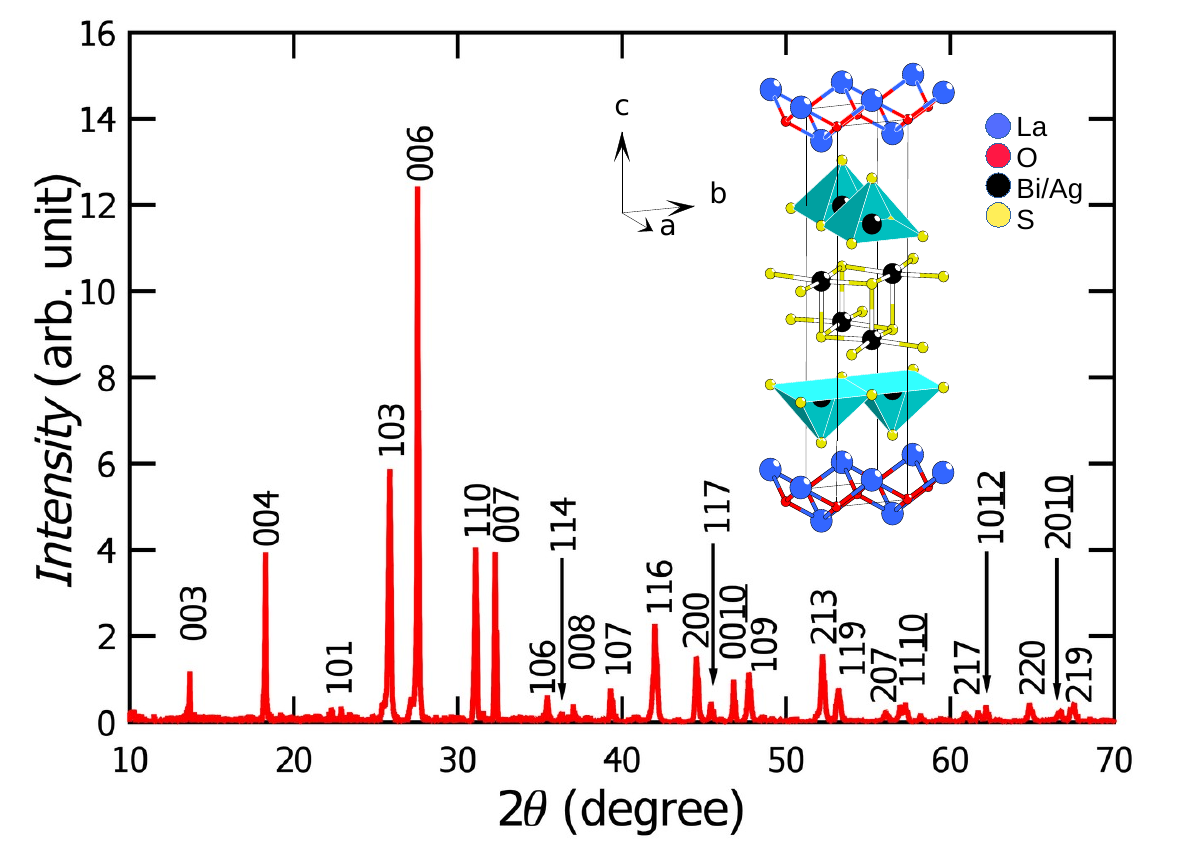}
	\caption{\label{fig1} Powder x-ray diffraction pattern of polycrystalline La$_2$O$_2$Bi$_3$AgS$_6$ at ambient pressure. The inset displays a schematic drawing of the crystal structure of La$_2$O$_2$Bi$_3$AgS$_6$. }
\end{figure}

\section{Results and discussion}

Figure \ref{fig1} shows the powder x-ray diffraction pattern of polycrystalline La$_2$O$_2$Bi$_3$AgS$_6$ at ambient pressure. The peaks observed in the pattern can be indexed to a tetragonal crystal structure with a space group of $P4/nmm$, which is consistent with previous reports \cite{Hijikata2017,Jha2018}. No major impurity phases are found in our data, indicating that our samples are of high purity. The inset in Fig. \ref{fig1} shows the schematic drawing of the crystal structure of La$_2$O$_2$Bi$_3$AgS$_6$. It consists of [La$_2$O$_2$] blocking layers and [M$_4$S$_6$] conducting layers stacking along the $c$-axis, where the ratio of Bi:Ag at the M site is 3:1. Comparing with the most studied BiS$_2$-based superconductor LaO$_{0.5}$F$_{0.5}$BiS$_2$ \cite{Mizuguchi2012a,Mizuguchi2014a}, both compounds share the similar blocking layers but the blocking layers in La$_2$O$_2$Bi$_3$AgS$_6$ become thicker. \par

Figure \ref{fig2} displays the ambient pressure data of La$_2$O$_2$Bi$_3$AgS$_6$ for Samples E1 and M1, which come from the same batch. At high temperatures ($>$200 K), both samples exhibit a metallic-like behavior. As temperature goes down, the resistivity starts to increase, demonstrating a semiconductor-like behavior, and shows a weak anomalous hump around 155~K for sample E1. At low temperatures, a superconducting transition is observed at $T_c \sim$1.0 K (inset of Fig. \ref{fig2}(a)), where $T_c$ is defined as the temperature corresponding to 90\% of the normal-state resistance. Our resistivity data are generally consistent with that observed in previous studies \cite{Hijikata2017,Jha2018}, in which the anomaly around $155$~K has been attributed to a CDW phase transition. \par

\begin{figure}[!tb]
    \centering
	\includegraphics[width=7cm]
	{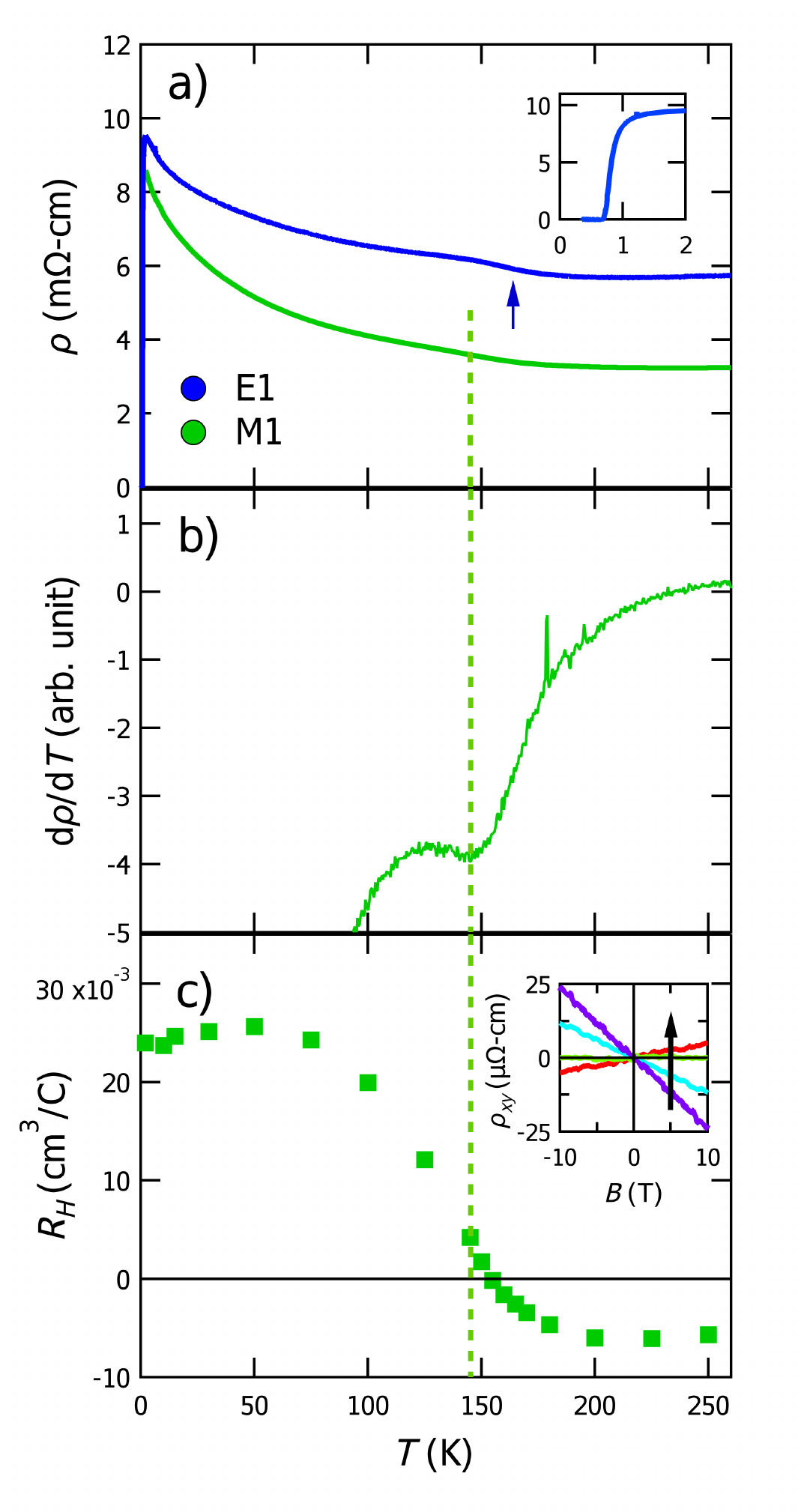}
	\caption{\label{fig2} Temperature dependence of (a) electrical resistivity of samples E1 and M1, (b) the first derivative of the electrical resistivity of Sample M1, and (c) the Hall resistance of Sample M1 at ambient pressure. The dashed line indicates the temperature $T^*$ at which the CDW-like transition occurs for Sample M1, whereas $T^*$ for Sample E1 is indicated by the vertical arrow. The inset of (a) zooms into the superconducting transition of Sample E1. The inset of (c) shows $\rho$$_x$$_y$$(B)$ at 2~K, 125~K, 155~K and 300~K, with the arrow indicating the direction of increasing temperature. 
	}
\end{figure}

To better visualize this anomaly, we also plot the first derivative of resistivity against temperature in Fig. \ref{fig2}(b), which shows a minimum at the anomaly. In this study, this minimum is used to define the transition temperature of the CDW-like transition $T^*$. The value of $T^*$ at ambient pressure was observed to be sample dependent in the range $\sim$155$\pm$8~K, which is marked by the dashed line in Fig. \ref{fig2} for Sample M1 and the vertical arrow in Fig. \ref{fig2}(a) for Sample E1. \par

In order to understand the origin of the anomaly around $T^*$, a Hall effect measurement was made on Sample M1. $\rho_{xy}$(B) was measured between $-10$~T and $10$~T as shown in the inset of Fig. 2(c). The raw resistivity data in all datasets were antisymmetrized to remove the $\rho_{xx}$ component. Hall coefficient at various  temperatures can be extracted by calculating the slopes of these $\rho_{xy}(B)$ curves, as shown in Fig. 2(c). Around room temperature, $R_H$ is negative and nearly temperature independent. Near $T^*$, $R_H$ undergoes a transition from negative to positive upon cooling, and the magnitude of $R_H$ eventually reaches a larger value than the room temperature value. Such a sign-changing $R_H(T)$ is interesting, and bear striking resemblance to the case of EuBiS$_2$F, which is another system exhibiting a possible CDW state \cite{Zhai2014}. At the level of a single-band approximation, the increase of $|R_H|$ at low temperatures reveals a decrease in carrier density, which can be understood as the possible opening of a CDW gap on the Fermi surface. \par

\begin{figure}[!tb]
    \centering
	\includegraphics[width=7cm]
	{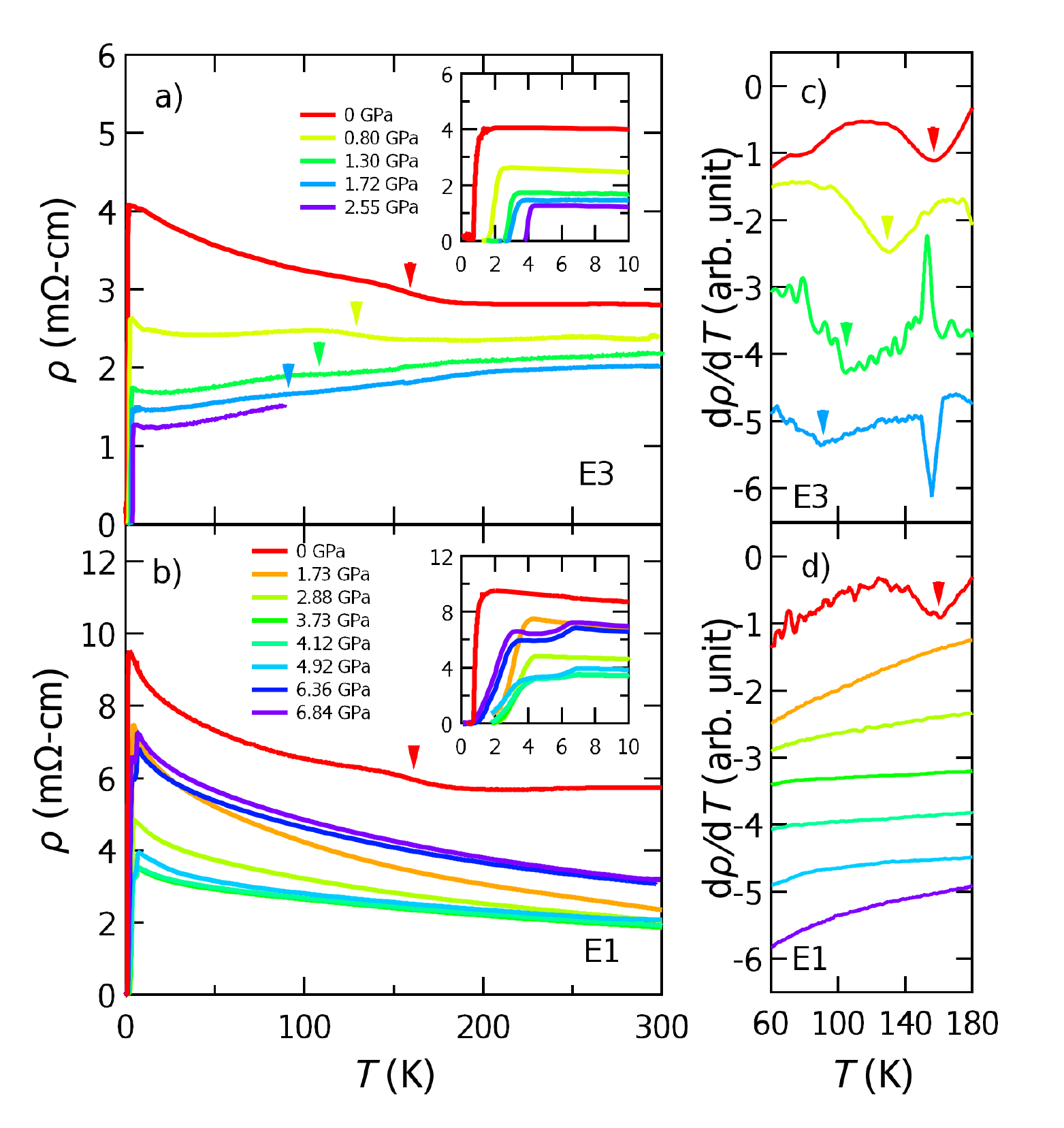}
	\caption{\label{fig3} Temperature dependence of the electrical resistivity of (a) Sample E3, and (b) Sample E1 at various pressures. The insets zoom into the superconducting transition. Vertical arrows mark $T^*$. The first derivative of resistivity with respect to temperature of (c) Sample E3, and (d) Sample E1 for selected pressure points. The color coding of (c) and (d) is the same as in (a) and (b), respectively. 
	}
\end{figure}

Figure \ref{fig3} shows the temperature dependence of resistivity measured for Samples E3 and E1 at different pressures. The ambient pressure data for Sample E1 is the same as that of Fig. \ref{fig1}(a). With an increasing pressure, the anomaly in resistivity for Sample E3 shifts to lower temperatures, and becomes less prominent. Using the temperature derivative of resistivity, we track the evolution of $T^*$ under pressure, and the corresponding values of $T^*$ at each pressure point are marked by the vertical arrows in Figs. \ref{fig3}~(a) and (c). The anomaly in resistivity for Sample E1 is much less prominent and cannot be determined accurately at high pressures. All datasets show superconductivity at low temperatures, which can be visualized in the insets of Figs.~\ref{fig3} (a) and (b). An anomalous drop in resistivity near 7~K was observed above 4~GPa, as displayed in the inset of Fig.~\ref{fig3}~(b). The origin of the drop is presently unknown. \par

The pressure dependencies of $T_c$ and $T^*$ for samples E1, E2 ($\rho(T)$ not shown) and E3 are summarized in a temperature-pressure phase diagram in Fig. \ref{fig4}. As pressure increases, $T^*$ of Sample E3 decreases linearly at a rate of 40.8~K/GPa up to 1.7~GPa. Using only the data of Sample E2 up to 2.1~GPa, d$T^*$/d$p$ is $-$41.9~K/GPa, close to the value obtained for Sample E3. At higher pressures, the $T^*$ anomaly} becomes too broad and can no longer be distinguished in the derivative data. In Sample E3, the anomaly can only be detected at ambient pressure. With all the $T^*$ data available, the overall d$T^*$/d$p$ is calculated to be $-$41.0~K/GPa. Linearly extrapolating $T^*$ to zero temperature with this rate, we find that $T^*$ reaches zero at $\sim$3.9~GPa. On the other hand, $T_c$ first increases with pressure and reaches a maximum of 4.1~K at 3.1~GPa. We point out that despite the usage of three samples from the same batch, and hence three sets of high pressure experimental data, $T_c(p)$ from these different samples shows a good agreement. Finally, we note an interesting anticorrelation between the pressure evolution of $T_c$ and the normal state resistivity, as benchmarked by $\rho_{10\rm{K}}$. Such a behaviour, accompanied by an analogous superconducting dome, and a change in sign of $R_H$ have also been observed in recently discovered nickelate based superconductor, where the electronic and superconducting properties can be tuned by varying the Sr content of Nd$_{1-x}$Sr$_x$NiO$_2$~\cite{Li2020,Zeng2020}. \par

\begin{figure}[!tb]
    \centering
	\includegraphics[width=8cm]
	{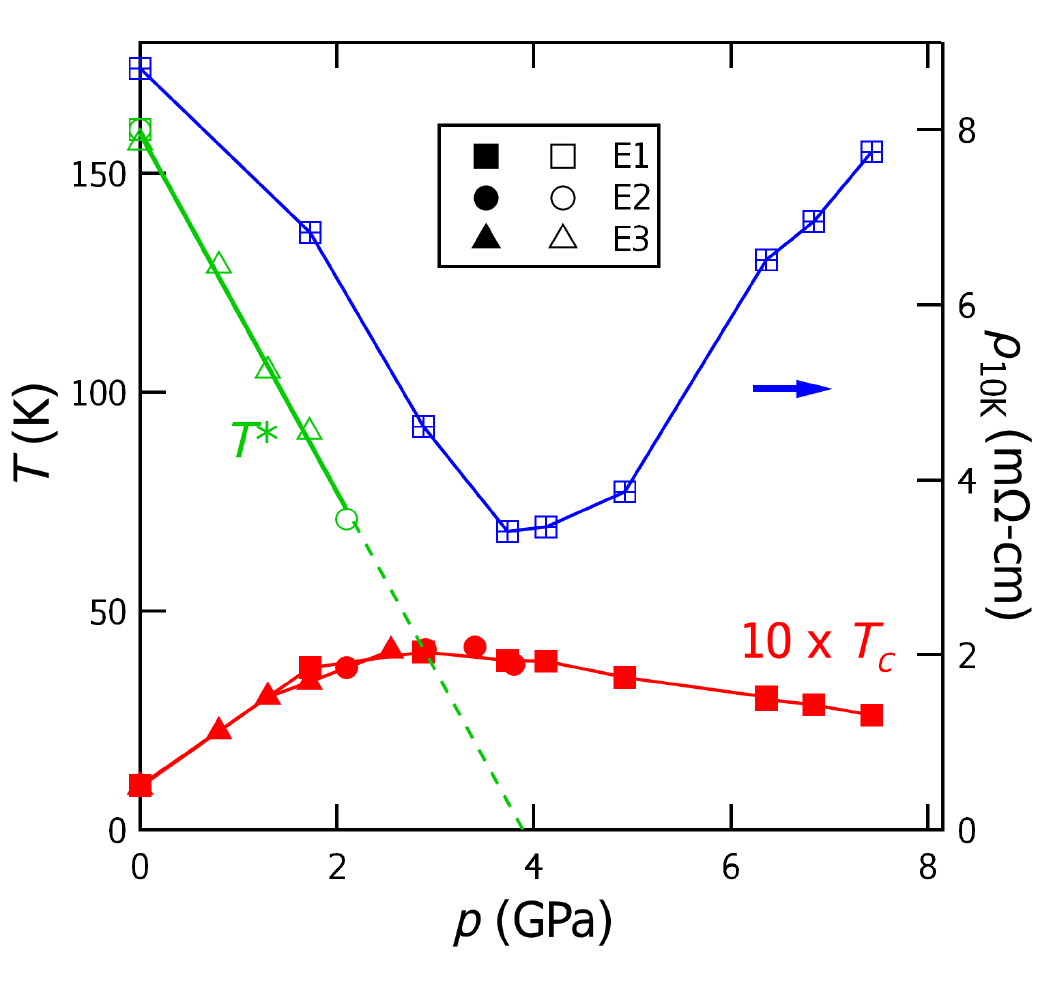}
	\caption{\label{fig4} Temperature-pressure phase diagram for samples E1, E2 and E3. The values of $T^*$ and $T_c$ are extracted from the resistivity data shown in Fig.~\ref{fig3}. The solid line is the linear fit to $T^*(p)$ while the dashed line shows how $T^*$ extrapolates to 0~K at 3.9~GPa. Blue markers represent resistivity at $10$~K for Sample E1.  
	}
\end{figure}

It is instructive to compare our temperature-pressure phase diagram with the phase diagram obtained via Sn substitution reported by Jha {\it et al.} \cite{Jha2018}. In terms of $T_c$, our phase diagram is qualitatively similar to the phase diagram constructed by Jha {\it et al.} However, $T^*$ appears to be much more sensitive to pressure than to the Sn concentration.  According to our phase diagram, a dome-like $T_c(p)$ where $T_c$ peaks near the pressure where $T^*\rightarrow 0$ 
hints at the importance of fluctuations related to the $T^*$ transition on superconductivity in La$_2$O$_2$Bi$_3$AgS$_6$.
Thus, our phase diagram resembles the phase diagrams of other superconductors, including heavy fermions~\cite{Gegenwart2008}, (Sr,Ca)$_3$(Ir,Rh)$_4$Sn$_{13}$~\cite{Klintberg2012,Goh2015,Yu2015}, and iron-based superconductors~\cite{Chen2009,Pratt2009,Parker2010,Pratt2011,Luo2012} where superconductivity is optimized in the vicinity of a putative quantum phase transition. If $T^*$ can be firmly established to be a transition temperature of a CDW state, La$_2$O$_2$Bi$_3$AgS$_6$ will become another good platform to study the interplay between CDW and superconductivity. Further investigations are desirable to uncover the nature of the phase below $T^*$ in La$_2$O$_2$Bi$_3$AgS$_6$, and to understand how the ordered state below $T^*$ interacts with superconductivity. \par

\section{Summary}

In summary, we have measured the electrical resistivity of La$_2$O$_2$Bi$_3$AgS$_6$ and constructed the pressure-temperature phase diagram up to 7.43~GPa. A resistivity anomaly at $T^*$, which is possibly related to a CDW phase transition, decreases linearly with pressure and can be extrapolated to 0~K at 3.9~GPa. $T_c$ forms a broad dome, reaching a maximum of 4.1~K at 3.1~GPa, which demonstrates a strong interplay of between the CDW-like transition and superconductivity.\par

\ack
 This work is supported by Research Grants Council of Hong Kong (CUHK 14300418, CUHK 14300117), CUHK Direct Grants (No. 4053345, No. 4053299, No. 4053463), CityU Start-up  Grant  (No.  9610438).

\section*{References}
\bibliographystyle{iopart-num}
\bibliography{main}

\end{document}